**Title**: From F=ma to Flying Squirrels: Curricular Change in an Introductory Physics Course

**Authors**:

   Brian O'Shea[*], Lyman Briggs College and the Department of Physics and Astronomy, Michigan State University

   Laura Terry, College of Education, Michigan State University

   Walter Benenson, Lyman Briggs College, Michigan State University

**Abstract**:   We present outcomes from curricular changes made to an introductory calculus-based physics course whose audience is primarily life science majors, the majority of whom plan to pursue post-baccalaureate studies in medical and scientific fields.  During the 2011-12 academic year, we implemented a "Physics of the life sciences" curriculum centered on a draft textbook that takes a novel approach to teaching physics to life science majors.  In addition, substantial revisions were made to the homework and hands-on components of the course to emphasize the relationship between physics and the life sciences and to help the students to learn to apply physical intuition to life science-oriented problems.  Student learning and attitudinal outcomes were assessed both quantitatively, using standard physics education research instruments, and qualitatively, using student surveys and a series of post-semester interviews.  Students experienced high conceptual learning gains, comparable to other active learning-based physics courses.  Qualitatively, a substantial fraction of interviewed students reported an increased interest in physics relative to the beginning of the semester.  Furthermore, more than half of students self-reported that they could now relate physics topics to their majors and future careers, with interviewed subjects demonstrating a high level of ability to come up with examples of how physics affects living organisms and how it helped them to better understand content presented in courses in their major.

* Corresponding author: Brian O'Shea, oshea@msu.edu

## Section 1: Introduction

Experienced practitioners of medicine and the life sciences are keenly aware of the role that physics plays in the structure and behavior of living systems at all levels, from molecules to ecosystems.  Beginning students of these subjects, however, have a very difficult time connecting the topics that they learn in a typical introductory physics course sequence to their chosen college major and career path – a source of great frustration to their instructors!  This disconnect is only becoming more important as life scientists rely more heavily on tools, both experimental and theoretical, that require a strong quantitative and physics background to understand and use effectively.  In response to this, recent reports have called for curricular changes that more fully integrate the physical sciences into the education of life science students, and of those who want to pursue careers in medical fields (AAAS 2009; AAMC-HHMI 2009; National Research Council 2010).  One specific example of this is the document "Vision and Change in Undergraduate Biology Education: A call to Action," published by the American Academy of Arts and Sciences (AAAS 2009), which argues for a set of core concepts that all undergraduate biology students should understand, including  "Structure and Function" and "Pathways and transformations of energy and matter."  The former concept requires students to understand the physics behind the functioning of structural components of living organisms, and the latter concept requires students to understand how the laws of thermodynamics govern the flow of energy in living systems.  More broadly, the Vision and Change document outlines a set of core competencies that all biology students need to develop, which include the "ability to use quantitative reasoning," the "ability to use modeling and simulation," and the "ability to tap into the interdisciplinary nature of science"  (AAAS 2009, pp. 14-15). The first two competencies tap directly into the core strengths of physics as a discipline – namely, its quantitative, model-oriented nature - and the latter offers an opportunity to more directly tie together the physical and life sciences.  To this end, some forays have been made toward the tighter integration of biological principles into physics courses, and vice versa – most notably by the NSF-funded, interdisciplinary NEXUS program at the University of Maryland (see Redish 2012) – however, these projects are in their infancy.

In this paper, we present outcomes from redesigning an introductory physics course sequence at Michigan State University to directly address the needs of life science students.  During the 2011-2012 academic year, we implemented a "Physics of the life sciences" curriculum in a physics course sequence whose population is primarily made up of undergraduates majoring in the life sciences.  This curriculum is based on a draft textbook that takes a novel approach to teaching physics that focuses on the way in which fundamental physical principles dictate the form and function of organisms.  In addition to a new textbook, substantial revisions were made to the homework, recitation, and laboratory components of the course to emphasize the relationship between physics and the life sciences and to help the students learn to apply physical intuition to life science-oriented problems.   We primarily discuss changes made in the first semester of the course sequence, where a broader array of both quantitative and qualitative assessments are available to analyze than in the second semester of the sequence. In this paper, we attempt to address the question "Can we measurably and positively change conceptual understanding and attitudinal outcomes of our student population by using discipline-specific physics examples?"

This paper is organized as follows. In section 2, we discuss the context within which these curricular changes were made and the population of students taking the course. Section 3 describes the new curriculum that has been implemented. Section 4 describes the means we used to assess the changes to the first semester of the course sequence, with the results being described in Section 5. Finally, in Section 6, we discuss the results and some further changes that we will make in response to these results.

## Section 2: Context of the curricular changes

The course used in this project is the first semester of an introductory calculus-based physics course taught in Lyman Briggs College (LBC), a residential college within Michigan State University (MSU). Lyman Briggs College is one of MSU's flagship programs, and is an undergraduate living-learning community whose main goal is to bridge the sciences and humanities through interdisciplinary teaching and learning (Lyman Briggs College 2012). The faculty of Lyman Briggs College is composed of scientists, mathematicians, and humanities scholars whose work focuses on the history, philosophy, and sociology of science. Faculty members in Lyman Briggs College often participate in interdisciplinary teaching and research, and encourage their students to do the same. This is done most visibly through the NSF-funded Bridging the Disciplines with Authentic Inquiry and Discourse program (BRAID; Luckie et al. 2012). The first iteration of this program attempted to explicitly make connections between LBC-taught first year courses – specifically, biology and chemistry; later incarnations include other courses and disciplines. (Luckie 2012)

Lyman Briggs College typically only accepts students as freshman, and takes approximately 625 students per year on a first-come, first-served basis. Enrollment into the college occurs when students accept an offer from Michigan State University and declare LBC to be their college of choice. After enrolling in LBC, students can choose any number of majors within the college, including the sciences, mathematics, computer science, and the history, philosophy, and sociology of science (HPS). All students in the college are required to take a core set of classes, including two-semester introductory biology, chemistry, and physics sequences, as well as calculus and several HPS courses. Students who decide they wish to major in other disciplines, or to pursue the equivalent non-LBC science major, may opt to transfer to another college within MSU if they choose. Although a range of disciplines are available to Lyman Briggs students, the majority of students who graduate from the college major in the life sciences (85% in 2011), and more than 80% of students in a given graduating class plan to pursue post-baccalaureate education of some kind, with popular choices being medical school, veterinary school, and graduate school in medical and life science-related fields.  In the 2010-11 academic year, the composition of the incoming class of 627 students was 55% female, 22% minority students, and 8% out-of-state students (with no distinction made in this category between domestic and foreign students).  Due to attrition, the total size of the college in the 2010-11 academic year was 1,865 students, with the most recent graduating class being composed of 293 students, 55% of whom were female.  While attrition from Lyman Briggs College is substantial (more than 50% of students starting in Lyman Briggs do not finish their degree in the college), the vast majority of students beginning at Lyman Briggs remain at the university and complete their degrees – roughly 90% of students who start in LBC as freshmen end up graduating from Michigan State University within six years.

Most colleges and universities typically offer two introductory physics course sequences: a three-semester calculus-based sequence targeted at engineers and physical scientists, and a two-semester algebra-based sequence for all other students who are required to take physics.  Michigan State University offers these two options to all students; however, Lyman Briggs College offers a third option.  This third option is a two-semester, calculus-based physics sequence (LB 273 and LB 274) that is only offered to LBC students.  Historically, this course sequence covered approximately the same material as a standard two-semester algebra-based course (kinematics, dynamics and a small amount of thermodynamics in the first semester, and electricity and magnetism, optics, and 'modern physics' in the second semester), albeit in somewhat greater depth.  This course sequence has three 50-minute lectures per week, as well as one three-hour "hands-on session," which is a combination of recitation and laboratory activities with content that is integrated tightly with the lectures and reading material.  The lecture sections range from 50-130 students, and the hands-on sessions have up to 24 students, with students working in either pairs or groups of four under the supervision of two undergraduate Learning Assistants, who are older LBC students that took the course in previous years (see Otero et al. 2010 for an example of a similar program).  A variety of active learning techniques are used in both the lecture and hands-on sections, including think-pair-share conceptual and quantitative questions (King 1993) and individual and small-group problem solving in lecture, and concept-focused Washington Group physics tutorials (McDermott et al. 2003), context-rich physics problems (Heller 2012), and inquiry-driven laboratory exercises in the hands-on sessions.  Pre-class questions on the reading and weekly homework is assigned and graded using the LON-CAPA course management system (Kortemeyer et al. 2008).

Lyman Briggs students have the option of either taking the two-semester calculus-based sequence offered by Lyman Briggs College, or enrolling in one of the two university-wide, and much more traditional, introductory physics course sequences.  Over the course of several semesters, a steady decrease in enrollment in the LBC physics course sequence was observed, with the majority of these students choosing to take MSU's algebra-based introductory physics course sequence instead. In an attempt to understand this phenomenon, LBC faculty asked Lyman Briggs students who had taken one of the three physics course sequences to fill out an anonymous survey explaining their choice of courses. Overwhelmingly, it was reported that the Lyman Briggs physics sequence was more rigorous and required more effort of the students than the university's algebra-based sequence, and that the acquisition of the highest possible grade was considered to be much less certain. Furthermore, many students expressed a lack of understanding as to why they should have to take physics, and how it might help them in the future. Taken together, this caused many LBC students to opt to take the course that was perceived as being the least amount of work, with the highest certainty of getting the best possible grade.

## Section 3: The New Curriculum

In response to the challenges described in the previous section – namely, falling course enrollment and students' lack of understanding of the relevance of physics to the life sciences and their career plans – the Lyman Briggs College physics faculty decided to more explicitly make connections between physics and the life sciences. After considering the options, we settled on the draft textbook Physics of the Life Sciences, by Prof. Timothy McKay (University of Michigan). McKay's textbook is different from other nominally similar texts (such as Newman 2008) in several intriguing ways. This textbook takes a novel approach to teaching physics to life science majors, by explicitly focusing on how physical principles dictate the shape, size, and structure of organisms, with an emphasis on physical scaling laws such as the relationship between an object's mass and its surface area. Such scaling laws are touched upon repeatedly to explain such disparate phenomena as the systematic way that metabolic rates vary with animal size, the convergent evolution that is observed in flying creatures in different animal kingdoms and how the behavior of diffusion at the atomic level dictates the size of single-celled organisms and the requirement that large animals have circulatory systems in addition to lungs or gills. In addition, McKay's textbook spends substantially more time on thermodynamics, statistical mechanics, and fluid statics and dynamics than a standard introductory physics textbook, and less time on kinematics and dynamics – changes that reflect the needs of life science students. Finally, McKay's book is calculus-based, whereas the standard physics texts targeted toward life science majors are algebra-based (e.g., Newman 2008), and thus lack the mathematical rigor that we desire, or are true biophysics textbooks, which are aimed at much more advanced students who have already taken basic physics and have a much more extensive background in mathematics.

Other aspects of the course were also modified in support of this new curriculum. The lecture sessions, which were already highly interactive, were modified to include both conceptual and quantitative questions with a life science theme, with additional problems explicitly designed to introduce students to the idea of modeling living systems. An example problem given in lecture, used during the section of the course dedicated to understanding forces, is as follows (with the accompanying illustration being shown in Figure 1):

> *Consider a person holding a dumbbell in their hand and performing a bicep curl. When doing this, they hold the dumbbell by their side, and then, without moving their elbow, slowly lift the weight until it is touching their chest. Consider the diagram shown in Figure 1, which shows the position of the weight, bones, and bicep when the person's forearm is being held horizontal. Given the numbers provided, how much force must this person's bicep muscle exert at this point in the curl?*

Problems of this sort require a multi-step solution, and require the understanding of several physics tools and concepts, including free body diagrams, force, and torque. They also tie into a broader discussion of locomotion – muscles only pull, and cannot push, so the construction of both vertebrates and invertebrates with exoskeletons are designed around this fundamental constraint.

We also developed a substantial number of new homework problems within the LON-CAPA course management system, which explicitly use the physics principles discussed in lecture and in the reading materials to understand a wide array of situations that may be of interest to life science students, while striving to ensure that these problems are authentic to the disciplines involved (e.g., Watkins et al. 2012) and emphasize the conceptual thinking that the students will find most useful in their careers (e.g., Nichparenko 1985). For example, problems created for the first semester of the course sequence explore the following topics:

- How the response of healthy bones and osteoporotic bones to pressure differs, using experimental data from Thurner et al. (2005), to understand why people with arthritis are more prone to breaking their bones when falling than are healthy individuals.
- How the shape and size of animals result in different free-fall velocities (e.g., why an elephant has a higher free-fall velocity than a mouse, and thus why mice are known to fall from great distances without harm).
- How the diffusion rate of the adenosine triphosphate (ATP) molecule within the body sets fundamental limits on the duration of high-intensity exercise.
- The relative ability of organisms of different sizes to generate heat and dissipate it (and thus why small animals tend to have problems with maintaining a high enough body temperature, and large animals are prone to overheating).
- The relative blood pressure in wider areas of a blood vessel (i.e., aneurysms) and narrower areas of a blood vessel compared to the mean, and the effects that this change in pressure may have on the stability of the circulatory system (thus explaining why aneurysms and blocked arteries are of such grave concern).

Similar problems were developed for our hands-on sessions, where students were asked to work in small groups to model, e.g., the forces and torques on the human arm when lifting a heavy object, and on both the midterm and final exams, which were entirely free-response. A representative free-response exam question from the first midterm exam is:

> *Imagine that your evil physics professor lured you into a machine that shrunk you to be 1 cm tall (isomorphically, of course).*
>
> *A) After a bout of maniacal laughter, he got distracted and left you on top of a standard-height table. Would you be able to safely jump from the table to the floor? Why or why not?*
>
> *B) In your shrunken state, you adopt a carpenter ant as a means of transportation. You manage to get back into the machine, which transforms you back to your normal size. As a side effect, your trusty steed is now almost six feet tall! What is at least one scaling-related challenge that a human-sized carpenter ant would face?*

Correctly solving this exam question requires students to couple their understanding of the scaling law that controls the relationship between an object's mass, surface area, and the cross-sectional area of its limbs. These factors dictate the relative importance of forces such as gravity and wind resistance, as well as the ability of a creature's limbs to support its body.

## Section 4: Assessment of Changes

In this section and the next, we focus on assessment of, and outcomes from, only the first course in the two-semester sequence, where a broader array of both quantitative and qualitative assessments are available to us. The reasons for this are primarily practical: interviews focusing on the first semester of the course sequence (which was offered in the fall) were conducted over winter break or at the start of the spring term, when the course was still fresh in the students' minds. This was effectively impossible to do at the end of the spring term.

The changes made to this course were assessed in several ways. We administered two standard physics education research instruments – the Force Concept Inventory (FCI; Halloun & Hestenes 1985a, 1985b) and Maryland Physics Expectations Survey (MPEX; Redish et al. 1997, 1998) – as both pre- and post-tests during the first semester of the course sequence. The Force Concept Inventory is a 28-question multiple-choice test that is used to assess students' conceptual understanding of Newton's Laws, and is widely used in high school and college-level mechanics courses (Hake 1998).  The Maryland Physics Expectations Survey is an attitudinal survey that assesses students' understanding of physics as a discipline and their expectations for, and attitudes about, taking a physics course by asking a series of questions with Likert-scale responses.   The MPEX has also been administered to a large number of PhD-possessing physicists, thus allowing the quantification of 'expert-like' attitudes.  Both the FCI and MPEX are standard and widely-used physics education research instruments, which allow us to quantitatively compare learning gains and student attitudes both to previous iterations of our own courses (where both instruments have been administered for several consecutive years) and with courses at other universities.  Students were given full participation credit for taking both the FCI and MPEX pre-tests, with no penalty on the FCI for incorrect answers, nor rewards for correct ones.  Given that the MPEX is an attitudinal survey, with no "correct" answer, this issue is not relevant.  Students received participation credit for both the FCI and MPEX post-tests, though with the FCI they were given a small amount of extra credit if their score was above 70%.  While in principle this might bias our results, this reward has been given every semester that the FCI has been administered in Lyman Briggs College, which reduces the chance of it affecting our comparisons in a non-systematic way.

Qualitative outcomes were assessed in several ways.  An anonymous mid-semester survey was administered during the seventh week of class.  This survey was open-ended, and asked questions such as "What's going well in the class?" and "What needs improvement?"   During the last week of class, students were asked to fill out a paper-based version of the Student Assessment of Learning Gains (SALG; Seymour et al. 2000), which in addition to the standard complement of multiple-choice questions asked for open-ended feedback on the course.  Students were prompted to think about different aspects of the course with five questions:

1. Which topics did you like the most, and which the least?
2. Which topics seemed most relevant to you, and why?
3. Which aspects of the course structure (lecture, hands-on sections, online homework, help room, etc.) were most and least useful to you, and why?
4. Do you have any other suggestions as to how we might improve the course?
5. Do you have any suggestions as to how I might improve my teaching?

 Students were not given any external incentive to complete either of these surveys, though we note that the SALG was administered during class time to enhance the response rate.  Also, students are strongly encouraged to give extensive written feedback in all Lyman Briggs Courses, so response rates are typically high - 71 students out of 77 (92%) filled out the multiple choice component of the end-of-semester evaluation, and 61 of those (86%) provided substantial written feedback as well).

In addition to written surveys, a total of fifteen students were interviewed by one of the co-authors of this paper (Terry), and anonymized transcripts of these interviews were provided to the primary author (O'Shea) for analysis (which is described in more detail in section 5.2).  The interviewer (Terry) has no association with the class outside of the interview process, and students were informed both in the invitation email and prior to the interview that course staff (in particular, the primary author of this paper and primary course instructor, O'Shea) would know whether or not they participated in the interviews.  Of the fifteen interviewees, 9 were selected at random from the course roster, with three students apiece chosen from the top, middle, and bottom third of the course grade distribution.  The remaining six students were randomly selected from a pool of sixteen students suggested by the course faculty as being particularly likely to have insightful thoughts on the physics course that they had just completed.  The interviews were conducted between the end of the fall semester final exam session and the second week of the spring semester, and were done using an interview protocol that combined a pre-generated set of questions with in-depth followup questions to probe student responses more fully.  The pre-generated questions targeted student motivations ("Why did you take this class?"), connections that the students made to course content  ("What did you particularly connect to in this course?  Were there any specifically memorable assignments, discussions, or lectures?"), and impact of the course's learning goals on the student's thinking ("How did the topics discussed during the semester tie together?") and on their life outside of their physics course ("Did you talk about things that were covered in class with other people who were not enrolled in the course?  If so, could you give me an example?").  In addition to pre-generated questions, the interviewer followed up on students' comments to get more detail or to clarify vague statements.  The interview transcripts were examined by the primary author (O'Shea) for common themes, and for the correlation of student responses with other factors, such as the student's grade in the course and their major.

## Section 5:  Results

In Fall 2011, LB 273 (Physics I) was co-taught by two of the authors of this article (O'Shea and Benenson), with Brian O'Shea running the lecture and LON-CAPA components of the course, and Walter Benenson running the hands-on component. 75 students were enrolled in the course at the beginning of the Fall semester, and 77 students completed the course.  The increased number over the semester is due to a handful of students switching from the university's algebra-based course into the Lyman Briggs physics course during the first week of the semester, and a few students dropping the course.  The course was composed of approximately 54% women and 46% men, with the vast majority of students (73 of the 77 who finished the semester) self-identifying as life science majors.

## 5.1 Conceptual outcomes

Quantitatively, student outcomes paint a positive picture. The pre-test mean Force Concept Inventory score was 42.3% +/- 4.88% (N=75), which is substantially higher than previous semesters' pre-test mean score of 33-36% (+/- 2%) (Cruz et al. 2010). Given that the Lyman Briggs College student population has not evolved substantially over the last few years (in terms of demographics and standardized test scores), this argues that the students choosing to take the course in Fall 2011 were, on average, of higher quality than average. We speculate that this is due to a combination of two factors: less-prepared students systematically choosing to take alternate physics courses that are perceived to be easier (as suggested by the results of our anonymous survey from the summer before the course was offered), and better-prepared and more intellectually curious students systematically choosing to take the course specifically due to the advertised "Physics of the life sciences" focus. The mean post-test FCI score was 61.6% +/- 7.02% (N=77), compared to ~55% (+/- 3.5%) in earlier semesters (Cruz et al. 2010). The standard method to measure gain in such situations is to consider the *normalized gain*, $<g>$. This is defined as the improvement in students' understanding of Newton's Laws compared to the maximum possible gain available to them, or:

$<g>$ = (post-pre)/(100-pre)

where "post" is the post-test score and "pre" is the pre-test score. Our normalized gain is $<g>$ = 0.334 +/- 0.0381, which is statistically indistinguishable from previous implementations of this course (Cruz et al. 2010) and in line with introductory physics courses at other universities that use research-based interactive engagement techniques, and substantially higher than traditional, lecture-based courses (Hake 1998; Caballero et al. 2012). Overall, our students' performance on the Force Concept Inventory indicates that implementing the new curriculum has done no harm to our students' conceptual understanding of physics compared to the previous curriculum, although we have not substantially increased their conceptual understanding of Newton's Laws compared to previous semesters. Given that the goal of our curricular change was to more explicitly make connections between physics and the life sciences, and not to increase conceptual understanding of Newtonian mechanics, this result is quite acceptable. In fact, given that somewhat less time is spent on Newton's Laws, having learning gains that are comparable to previous semesters is actually quite encouraging!

In addition to the FCI, we administered the Maryland Physics Expectations Survey (MPEX) as both a pre-test and post-test. Prior work shows that it is typical for students' attitudes toward physics to become substantially more negative after a semester of introductory physics, although this strongly depends on the student population and course in question (Kortemeyer 2007; Redish 1997). In a typical analysis of the MPEX, questions are grouped into clusters that correspond to student attitudes about independence of learning, understanding of physics as a coherent, consistent framework, conceptual understanding of physics, links to reality, mathematics as a way of representing physical phenomena, and effort exerted toward the class. Furthermore, student surveys are compared to a set of "expert" answers, which were determined by the MPEX authors by administering the survey to a group of university physics faculty who were committed to reforming their teaching to make it more effective. This "expert" group provided answers that were strongly consistent (typically at the 90% level or grater), suggesting that mastery of physics is tied to having or developing the attitudes expressed. Table 1 shows the fraction of students giving favorable answers to the six clusters of questions, both before and after the semester (where "favorable" is defined as an answer on the same side of the Likert scale as that of the expert physicists). No statistically meaningful change was observed when comparing MPEX pre- and post-test values for five of the six clusters. We contrast the observed lack of change in Fall 2011 with the MPEX pre- and post-test values observed in previous iterations of this course (taught in Fall 2009 and 2010), where student MPEX scores become substantially more negative in the "reality link" cluster, meaning that students' belief that the ideas learned in this course are widely applicable to reality actually *declined* in previous semesters (see Kortemeyer 2007). This is an interesting result – while the students enrolled in this class don't seem to be fundamentally changing their thinking about physics as a discipline and in its connection to 'reality,' the change in curriculum is not hurting (and in some sense is a positive change compared to the previous curriculum). We note that we do see a substantial negative change in the "effort" cluster, from a favorability fraction of 0.675 before the semester to 0.549 afterward. This is typical of the vast majority of physics courses, both traditional and reformed, and is believed to be due to the change from "pre-course optimism" to "post-course reality checks" (Redish 1997). We also note that our pre-course "favorable" ratings are somewhat higher than in previous semesters (see Kortemeyer 2007, Table II), which may indicate that the student population enrolled in the class has a more positive overall attitude toward physics (or our specific course sequence) than the student populations enrolled in previous iterations of the course. This may introduce a bias in the results of our survey.

A final quantitative indicator of the impact of the curricular changes we have implemented is year-over-year course enrollment in the Lyman Briggs College physics sequence. Prior to the Fall 2011 semester, course enrollment had experienced a steady decline, to a minimum of 77 students in Fall 2011. Nearly all of the students enrolled in Fall 2011 (74 of 77 students, or 96%) enrolled in the second semester of the course sequence during the Spring 2012, with an additional 11 students transferring to the Lyman Briggs physics course sequence from elsewhere (the transfer of such a large number of students into the course sequence is unprecedented – typically, enrollment drops by roughly 5% between consecutive semesters of Physics I and Physics II in all of the physics course sequences at MSU). Of this population of 85 students, a total of 81 finished the course. In comparison, as of the beginning of the Fall 2012 semester, 124 students are enrolled in LB 273 – a 61% increase from the previous year's enrollment of the same class. It seems that the Lyman Briggs students are "voting with their feet," and may see the value of this curricular change. Of course, this is just one of many possible explanations for the observed increase. Other explanations of this outcome include, but are not limited to, the possibility that LBC students' perceptions of their options with regards to introductory physics courses may have changed from previous semesters, making the LBC physics sequence relatively more attractive. Further data are needed to determine the actual impact of the course revisions on student enrollment.

## 5.2 Affective outcomes

Results from our qualitative assessments were quite illuminating. As described in the previous section, Lyman Briggs College uses a paper-based version of the Student Assessment of Learning Gains (SALG) as an end-of-term course evaluation instrument, and a total of 61 students out of 77 finishing the course (79%) gave anonymous, open-ended feedback prompted by a series of questions, including "Which topics did you like the most, and which the least? Which topics seemed most relevant to you, and why?" (See Section 4 for a complete listing of questions.) No prompting with regards to the life science focus of the course was given.

The SALG free-response section and the interviews were examined for themes using the following methodology. First, the primary author of this paper (O'Shea) read through all of the SALG responses and interviews twice, shuffling the order between each reading. During this first cursory examination, several broad themes were tentatively identified for each set of documents, and a more detailed examination of each SALG form and interview was then conducted, using these themes as a template. We note that two separate sets of themes were used: no attempt was made to ensure consistency between the SALG written responses and the interviews. During this re-examination, it became clear that the tentative themes found were too highly articulated, and that what originally appeared to be different categories of comments were really the same sort of comment expressed in somewhat different ways (for example, "I absolutely hated the course pack because..." (followed by a list of reasons) and "The course pack could be improved" were originally counted as separate categories, but were later judged to be the same thematic issue, but expressed with different levels of substance). After the re-examination was conducted, a final pass through the data was made to double-check the categorization. One person was responsible for all categorization, and thus no reliability of categorization was possible (though some assistance was requested from colleagues for various categorizations, this was not done in a systematic way).

Examination of the SALG free-response items showed that several themes were present, with the two primary ones being 1) ability to connect physics to a life science-related topic, and 2) dissatisfaction with some aspect of the course pack used for the course. These themes were analyzed more closely, to identify substantial comments and also minor/incidental comments. 25 of the 61 students who provided written feedback specifically and positively mentioned the life science focus of the course, with 18 making in-depth comments such as "I really like the scaling law stuff... it helped me understand why you can have whales in the ocean but not on land!" and "This seems like it's going to be super-useful in the future when I take physiology." (anonymous, personal communications, Dec. 2012). A further 7 students made minor positive comments such as "I like how the examples relate to my major" or "I liked the life sciences examples." The majority of students interviewed (8 of 15) enjoyed the focus of the course, with a substantial number (4 students) indicating that despite early skepticism, they found that the course was ultimately quite interesting. Six students mentioned that they saw connections between the course, their own interests, and other scientific disciplines such as chemistry and biology. "Dr. Timothy McKay, from University of Michigan, he said in the beginning of the course that physics is the science that would explain chemistry, that would explain biology… When he talks about the thermodynamics… I did feel like the things I was learning from chemistry were really integrated into the physics, and physics. Learning what he taught me actually gave me a deeper understanding of that" (anonymous, personal communication, January 10, 2012). The interviewed students clearly liked the focus of the course, and thought that the life science-focused examples provided by the textbook, in lecture, and in the hands-on sessions helped to keep their interest. In general, the interview subjects liked the "big picture" aspect of the course – namely, using the laws of physics to understand the structure and behavior of organisms – and found that to be very memorable and relatable to their interests. Simply liking something does not imply understanding, of course – however, when asked to provide specific and physically correct examples connecting the physics they learned to their interests, 9 of the 15 students interviewed were able to do so. "Suspensory ligaments and how far they can stretch, going back to springs, how far they can stretch before they break. That was memorable to me. As an animal science major, when I come up with a circumstance where an animal has come up lame or something, I always think of that because I was able to make the connection" (anonymous, personal communication, January 19, 2012). Two students interviewed, who self-identified as non-life-science majors, specifically mentioned the life science focus of the course in a negative light, however, complaining that there was too many life science examples. "We're not all pre-meds, you know?" (anonymous, personal communication, Dec. 2012).

A significant number of interviewees (4 of 15) mentioned that, due to their interest in the material, they talked about these examples with people who weren't enrolled in the course. "I have talked to my parents a little bit probably relating the blue whale stuff. Those things definitely excite me, the general concepts, like I said, the overall class of relating physics to life sciences" (anonymous, personal communication, January 20, 2012). A few students (3 of 15 interviewees) mentioned that in additional to speaking to others about physics related ideas, they found themselves thinking about physics concepts in every day situations, such as walking on ice, driving a car, or watching an animal run.

A large number of students expressed concern or displeasure with the draft textbook (24 of 61 students on the SALG, and 7 of 15 students in interviews) – however, additional comments (8 on the SALG form and 4 from the interviews) indicated that this was not because of the focus of the textbook, but rather because of the lack of practice problems beyond the assigned homework problems and also confusing explanations of some phenomena. "I didn't like the course pack. You had to find the information. There was just a lot of useless fluff but once you found the information the course pack did help, it just took two or three times of reading it to actually get there" (anonymous, personal communication, January 19, 2012).  In fact, the life science focus of the textbook was a significant highlight of the student feedback, with 4 of 15 interviewed students bringing up the life science theme of the textbook in a positive light and without prompting. "That's the one thing I liked about the textbook was in every chapter they tried to relate it to like a life science, like in biology or something. Just using diffusion and using that as far as like the pulmonary region in the lungs and stuff was effective for me" (anonymous, personal communication, January 18, 2012).  It seems clear from this feedback that the published version of this textbook, containing example problems and supplemental homework problems, would substantially help the students' acceptance of the text.

### 5.3 Instructor Outcomes

In addition to feedback relating to students enrolled in the course, it is perhaps worth considering the impact that the new curriculum has had on the course personnel, particularly the instructors (and the first author of this article).  When the decision was made to switch to this curriculum, it sparked extensive and positive discussions with other Lyman Briggs College science faculty about the changes that were going to be made, the examples used, and the possible connections that could be explicitly drawn between the LBC physics sequence and the other introductory courses taught in the college (including continuing discussions about how to integrate the physics courses into the Lyman Briggs BRAID curriculum).  Over the course of the semester, faculty from other disciplines took the time to observe the class, examine course materials, and provide valuable feedback about the prior knowledge of the Lyman Briggs College students who had taken their courses before enrolling in physics.  Furthermore, as part of the process of digesting the new textbook and creating life science-focused homework and exam problems, the first author of this paper had the opportunity to think about a range of subjects that he had not been exposed to in nearly two decades, and which have been quite personally satisfying to explore.  Finally, the clear enthusiasm of the students toward the life science-focused examples used in the class has been profoundly refreshing, and has led to a much higher rate of positive student-faculty interactions than in previous semesters.

## Section 6: Discussion and future work

Converting the Lyman Briggs College introductory physics sequence from a fairly standard two-semester physics curriculum to one that focuses on the needs of life science students has been a significant positive step. Students' conceptual understanding of Newtonian mechanics has not been negatively impacted, based on quantitative metrics such as the Force Concept Inventory. Related to this, the course sequence covers most of the same topics, and with the same level of rigor, as in previous iterations of the course, but now includes additional topics that were not previously covered. The material that has been removed (primarily relating to kinematics and rotational motion) is of limited utility to our target population, while the content that has been expanded – including thermodynamics and fluid dynamics – is much more relevant to our students' needs in future courses. Furthermore, the qualitative data from interviews and surveys show the majority of students enrolled in the class clearly enjoyed the new focus of the course sequence, and stated that it helped to maintain their interest over the course of the semester and to make connections between physics and their own majors. The significant year-over-year enrollment increase also suggests that students in Lyman Briggs College feel that the knowledge they have gained is worth the additional effort (with an alternate interpretation being that students' perceptions of alternative courses have changed in the favor of our course, thus driving a climb in an enrollment). Overall, the experiment was a success.

There are several ways in which this study is limited. Primarily, we caution that the results presented in this paper may be biased due to the student population participating in the course. As discussed in Section 5, the students enrolled in LB 273 in Fall 2011 have substantially higher pre-test scores on the Force Concept Inventory than previous semesters (~42% compared to 33-36%). This may indicate that the students in the Fall 2011 semester have a higher academic ability than previous student populations, which may positively influence the receptiveness to the new curriculum. Alternately, it may simply represent more prior experience with physics among our student population.

Other limitations include the limited period of data collection (a single semester) and the choice of this time period (the first time the course is taught, when student and instructor enthusiasm may be affected most strongly). The results may not generalize to the second semester course, which focuses on topics in electricity, magnetism, and modern physics, and which was not evaluated in a similar manner. Similarly, we have made no attempts in this course or in previous courses to systematically measure students' understanding of topics that are more central in the reformed course (such as thermodynamics and fluid flow), or core skills such as problem solving. This is due in part to time constraints, but primarily because no standardized assessment instruments exist for these topics and skills.

Adoption of this physics curriculum has been deemed successful in Lyman Briggs College, and will likely continue to be used in the future with the full support of the faculty and college administration. This curriculum could be easily adopted by other institutions, although we note that the order and emphasis of topics is substantially different from that of a more standard course, which could be uncomfortable for some faculty (particularly physics faculty who are uninterested in or uncomfortable with the life sciences), and may require significant re-thinking of hands-on instructional components and an investment in new lab and/or demonstration hardware, as well as an investment in time to create a substantial quantity of life science-focused recitation, homework, and exam problems (with these last three points likely being the biggest initial barrier to adoption). Implementation of this curriculum requires support at the faculty and departmental level, at the very least – particularly in large-enrollment courses with multiple instructors and a great deal of inter-section coordination, as is common at many large R1 institutions. Given that the list of topics is qualitatively similar to a more standard introductory physics curriculum, it seems unlikely that new courses would have to be created or that substantial negative impact on courses that have introductory physics as a prerequisite would be felt. The reforms discussed in this paper are relatively easy to insert into an existing curriculum without changing other courses – if both life science and physical science faculty were supportive, however, one could imagine implementing a much more comprehensive set of reforms, including enforcing a much more extensive set of course prerequisites, including chemistry and biology (e.g., see the University of Maryland's Project NEXUS – Redish 2012). In this circumstance, one of the largest barriers to adoption would be the discomfort and/or hostility that many physicists appear to feel when asked to teach about life science topics that they may not have personally engaged with in many years, and a similar set of feelings with life scientists who are asked to engage with physics topics. This initial discomfort can likely be mitigated by careful choice of a faculty cohort and by encouragement and support at the college and university level.

As with any curriculum that is being used for the first time, there is substantial room for improvement in the next iteration of the course:

- The draft textbook was generally well received, but needs to be supplemented with additional practice problems until a published textbook is available.
- Additional homework problems need to be developed – particularly multi-part problems that explore the physical principles behind specific living systems.
- Appropriate laboratory exercises need to be developed – at present, the recitation-like component of the hands-on section is strongly coupled with the new curriculum, but our laboratory exercises are relatively standard.
- The introductory chemistry and biology course sequences that are taught in Lyman Briggs College are already tightly coupled through the BRAID program, and it would be logical to more formally integrate the physics course sequence into this program.

Of the points listed above, the last one has the potential to have the most impact on students' long-term success. Projects of an interdisciplinary nature are becoming more prevalent in the life sciences and in medicine, and the ability to apply multiple conceptual lenses (e.g., disciplinary viewpoints) to a problem is a very useful skill set. Giving students the opportunity to develop such skills, perhaps by some sort of extended project, may help to further increase student involvement and learning.

## Section 7: Accessing materials

All materials created for this course, including lectures, homework and exam problems, hands-on session materials, solutions for all materials, and a guide containing implementation advice are available for no charge. Please email the author directly at oshea@msu.edu to receive copies of these materials.

## Section 8: Acknowledgments

The authors acknowledge support for this effort from Michigan State University's Office of Faculty and Organizational Development and Lyman Briggs College through the Lilly Teaching Fellowship. We thank Punya Mishra for his invaluable advice and guidance, and Alicia Alonzo, Deborah DeZure, Cori Fata-Hartley, Elizabeth Gire, Elizabeth Simmons, Kendra Spence Cheruvelil, and Steven Wolf for useful discussions. We are very grateful to Timothy McKay for generously providing us with his draft textbook and other course materials. Finally, we are grateful to two anonymous referees for providing suggestions that have greatly improved the quality of this manuscript.

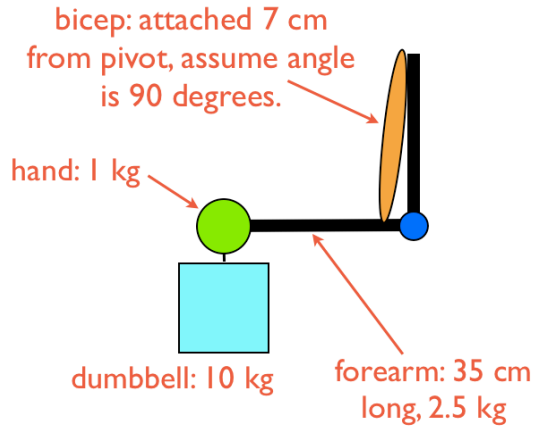

Modeling a bicep curl: how much force must your bicep muscle exert at this point in the curl?

bicep: attached 7 cm from pivot, assume angle is 90 degrees.

hand: 1 kg

dumbbell: 10 kg

forearm: 35 cm long, 2.5 kg

Figure 1: Diagram of a human arm lifting a dumbbell in a bicep curl. Students are asked to work in small groups to model the arm-dumbbell system at this point in the exercise, and must use several physics concepts (force, torque, and free-body diagrams) simultaneously.

| Cluster | Favorable Pre (N=69) | Favorable Post (N=75) | Shift [post-pre] |
|---|---|---|---|
| Independence | 0.571 +/- 0.066 | 0.541 +/- 0.062 | -0.030 |
| Coherence | 0.559 +/- 0.065 | 0.608 +/- 0.070 | +0.021 |
| Concepts | 0.457 +/- 0.053 | 0.478 +/- 0.055 | +0.021 |
| Reality link | 0.757 +/- 0.087 | 0.743 +/- 0.085 | -0.014 |
| Math link | 0.650 +/- 0.075 | 0.642 +/- 0.074 | -0.008 |
| Effort | 0.675 +/- 0.078 | 0.549 +/- 0.063 | -0.126 ** |

Table 1: Favorable fraction for the six dimensions of student expectations probed by the Maryland Physics Expectation Survey (MPEX), administered at the beginning (Pre) and end (Post) of the Fall 2011 semester, as well as the shift in scores (defined as "post – pre" score). Errors shown are standard errors. The "favorable fraction" is defined as the fraction of students whose answers to this cluster of questions are consistent with the answers given by a set of PhD-level university physics faculty who are interested in reforming their teaching and making it more effective. In the rightmost column, the only statistically significant shift (as determined by a one-way ANOVA test) is shown with a '**'.